\documentstyle[prd,aps]{revtex}
\input epsf
\begin{document}
\title{
Shell sources as a probe of relativistic effects in neutron 
star models}
\author{Zeferino Andrade and Richard H. Price}
\address{
Department of Physics, University of Utah, Salt Lake City, UT 84112}
\maketitle
\begin{abstract}
A perturbing shell is introduced as a device for studying the
excitation of fluid motions in relativistic stellar models.  We show
that this approach allows a reasonably clean separation of radiation
from the shell and from fluid motions in the star, and provides broad
flexibility in the location and timescale of perturbations driving the
fluid motions.  With this model we compare the relativistic and
Newtonian results for the generation of even parity gravitational
waves from constant density models.  
Our results suggest that relativistic effects will not be important
in computations of the gravitational emission  except possibly
in the case of excitation of the neutron star on very short time
scales.
\end{abstract}


\section{Introduction and overview}

For most astrophysical objects Newton's classical theory of gravity
gives a fully satisfactory description. Only when gravitational fields
become strong need one consider the possibility that general
relativistic effects may play a significant role. A standard index of
field strength is $GM/Rc^2$, where $M$ is an object's mass, and $R$ is
its characteristic size. This index is of order unity for black holes
and for the universe itself, and much smaller than unity for almost
all stars, galaxies, and other astronomical entities.  One exception
is neutron stars; for compact neutron stars $GM/Rc^2$ is on the order
of $~0.2$, and exotic equations of state could lead to even larger
values. Despite this, Newtonian gravity is used almost exclusively in
studying neutron stars. The obvious reason is the significant increase
in difficulty in giving a fully relativistic treatment of neutron star
structure, and the enormous increase of difficulty in dealing with
fullly relativistic dynamics, i.e.\,, with oscillations of neutron
stars.

Newtonian physics has been used even in studies (see for example
Refs.\cite{TURNER,WILL,BDLS}) of neutron stars as sources of
gravitational waves.  Of course, Newtonian gravity {\em per se} has no
gravitational waves, so the typical procedure is to compute
gravitational wave generation as a postprocessing step. More
specifically, Newtonian gravity is used to find the fluid motions
inside a neutron star associated with some event (core collapse in a
supernova, precession of a rotating neutron star, etc.).  Those fluid
motions are then used as sources in the ``quadrupole formula'' of
general relativity, just as known charge motions would be used in the
dipole formula of electromagnetism. If Newtonian theory predicts
periodic oscillations, the ``adiabatic approximation'' can be used to
give the damping of the oscillations due to gravitational wave
emission: the energy of the oscillations is taken to decrease at the
rate at which gravitational waves remove energy.

The sufficiency of this approximate procedure caused little worry
until relativistic modes of oscillation of neutron stars were
discovered \cite{KOJI,CHANDRA,KS1}  that had
no counterpart in Newtonian theory. The so-called $w$ modes are
qualitatively oscillations of the spacetime, like black hole
quasinormal (QN) modes, rather than oscillations of the neutron star
material like the $f$ (fluid) and $p$ (pressure) modes of the
Newtonian description. Partly due to the existence of these $w$ modes,
the general question of the sufficiency of Newtonian theory for
dealing with gravitational wave processes was emphasized by Andersson
and Kokkotas \cite{AK1}, and stirred considerable interest.

To give a clear answer to this question requires a specific and
astrophysically plausible event for which gravitational wave
generation can be computed in both Newtonian theory and fully
relativistically.  One would want, for instance, initial data for the
fluid and spacetime of a neutron star formed in a supernova core
collapse, but the possibililty of giving such initial data is at least
several years off. A more tractable model was needed.  Three separate
groups \cite{FGB,TSM,ZP} studied the problem of emission of
gravitational waves by a relativistic neutron star due to the close
passage of a perturbing particle. This model had the advantage of
definitiveness; there was no freedom in choosing (and biasing) the
initial spacetime perturbations to be particularly larger or smaller
than the fluid perturbations. This model allowed an investigation of
whether the excitation of $w$ modes was significant, but the model had
three serious shortcomings.  (i)~It was too restrictive. The only
parameters were the two constants specifying the particle orbit (say,
energy at infinity and impact parameter).  There could be significant
excitation only if the particle passed close to the neutron star, and
this constrained the perturbation to be neither very fast nor very
slow. Not only was the timescale limited, but it was coupled to the
choice of location of the perturbation. (ii)~It was difficult to
separate the radiation due to the neutron star (the radiation
of interest) from the radiation coming (in some sense) from the
orbiting particle. (iii) It was not clear how to compare the fully
relativistic computation from the Newtonian computation; at least
there was no attempt to do this.

We introduce here a very different model for investigating the
importance of relativistic effects in neutron stars, and possibly for
answering other questions.  We consider a perturbative spherical shell
around a neutron star at some radius $R_{\rm shell}$. In that shell we
dictate the time dependence of a multipole of surface mass-energy
density of the shell. There is no equation of state of the shell
material constraining our choice. The equations of motion of the shell
fix the surface stress once we have specified the surface energy
density, so picking that single function of time fully specifies the
source. In this manner we can independently choose where the
perturbation of the neutron star arises and what its timescale is.  We
can probe the response of the star to close and far perturbations with
slow motions or fast motions.  The shell probe has the nice feature
that it is straightforward to do the calculation in Newtonian theory,
so that a comparison can be made with the relativistic result.  A
further advantage is that both the Newtonian and relativistic
computations allow a separation of the gravitational waves from the
shell and those from the star. In the Newtonian computation, this is completely
straightforward. The waves from the star are those found from the
quadrupole formula applied to the motions only of the stellar
fluid. In the relativistic calculation the separation is only
approximate, and more care is needed.  From the star+shell results it
is necessary to subtract the radiation from the shell itself generated
in the background spacetime of the star. (The details of this
procedure will be given below.)

Our main purpose in the present paper is to present the method of
using a shell probe and to display its advantages.  For that reason we
limit the application of this method to the simplest model of a star,
the homogeneous incompressible perfect fluid (HIF) model. (See
\cite{AKK} for details about the QN modes of this model.) For this
model the only mode associated with motions of the stellar material is
a single $f$ mode. An important element of our results is that we will
present answers not only about excitation of $w$ modes, but about the
difference in the Newtonian and relativistic predictions of excitation
of the $f$ mode. In the interest of brevity we limit the analysis to
even parity perturbations. The excitation of $w$ modes for odd parity
should not be remarkably different from the excitation in even parity,
and odd parity motions do not couple to fluid motions.

The paper is organized as follows: The shell source model is
introduced in Sec.\,\ref{sec:shell}, and the equations governing even
parity perturbations due to the shell source are given in
Sec.\,\ref{sec:even}. Some details of the computational implementation
are given in Sec.\,\ref{sec:compimp} along with a discussion of the
method used for subtracting the shell contribution from the
relativistic calculation of the gravitational wave due to the shell
and star. Numerical results are presented and discussed in
Sec.\,\ref{sec:numres},  and conclusions are given in
Sec.\,\ref{sec:conc}.  Details of the Newtonian calculation are given
in the Appendix.  Throughout the paper we use geometric units $G=c=1$,
the metric signature $(-+++)$ and the conventions of Misner, Thorne
and Wheeler\cite{MTW}.
\section{Model: Perturbation of a static star by matter moving on a 
spherical shell}\label{sec:shell} 
We start with a static and spherically symmetric spacetime background metric
\begin{equation}\label{metric}
ds^{2}=-e^{\nu(r)}dt^{2}+e^{\lambda(r)}dr^{2}+r^{2}[d\theta^{2}+\sin^{2}\theta
d\varphi^{2}]\ ,
\end{equation}
describing both the interior and exterior of a star of a barotropic,
ideal fluid of mass M
and radius $r=R$. The stress energy of the fluid is
\begin{equation}
T_{\alpha\beta}=(\rho+p)u_{\alpha}u_{\beta}+pg_{\alpha\beta}\ ,
\end{equation}
where $\rho$ is the mass-energy density and $p=p(\rho)$ is the
pressure.  The mass function $m(r)$ is defined by
$e^{-\lambda(r)}=1-2m(r)/r$, and the structure of the stellar interior
is found by solving the hydrostatic equilibrium equations of general
relativity. (See, e.g.\,, Eq.\,(3) of \cite{ZP}.)  For simplicity, we
limit considerations to homogeneous incompressible fluid (HIF) stellar
models whose unperturbed interior metric is given by Eqs.\,(5) and (6)
of \cite{ZP}.  The exterior metric is simply the Schwarzschild metric,
with $m(r)$ equal to a constant $M$, and $\nu(r)=-\lambda(r)$. A
spherical thin shell of coordinate radius $r=R_{\rm shell}>R$
surrounds the star. We treat the shell as a perturbation of the
spacetime inside and outside the star and we analyze the perturbations
only to first order in the parameter of the perturbation. In this
order, the shell has spherical geometry described by the 3-metric,
\begin{equation}\label{shellmetric}
ds^{2}|_{\rm shell}=-\left(1-\frac{2M}{R_{\rm shell}}\right)dt^{2}
+R^{2}_{\rm shell}[d\theta^{2}+\sin^{2}\theta d\varphi^{2}]\ ,
\end{equation}
induced by the Schwarzschild metric. The metric of the perturbed
spacetime can be written as
\begin{equation}
g_{\alpha\beta}=g^{(0)}_{\alpha\beta}+h_{\alpha\beta}
\end{equation}
where the ``(0)'' index denotes the background solution, that of
Eq.\,(\ref{metric}). The Einstein equations to first order in
perturbations are
\begin{equation}\label{firstorder}
\delta G_{\alpha}\:^{\beta}=8\pi [\delta T_{\alpha}\:^{\beta}_{\rm fluid}+
\delta T_{\alpha}\:^{\beta}_{\rm shell}]\ .
\end{equation}
The perturbed stress energy has two contributions. One, denoted by
$\delta T_{\alpha}\:^{\beta}_{\rm fluid}$, is that of the fluid star
perturbations and is nonzero only inside the star.  The other, denoted
by $\delta T_{\alpha}\:^{\beta}_{\rm shell}$, is the stress energy of
the matter in the thin shell and is nonzero only outside the star. Its
form, in the coordinates of Eq.\,(\ref{metric}), is
\begin{equation}\label{shellstress}
\delta T^{\alpha\beta}_{\rm shell}=\sqrt{1-2M/r}\;S^{\alpha\beta}
\delta(r-R_{\rm shell})\ ,
\end{equation}
where
\begin{equation}
S^{\alpha\beta}=\lim_{\epsilon\rightarrow 0}\int_{R_{\rm
shell}-\epsilon}^ {R_{\rm shell}+\epsilon}
T^{\alpha\beta}\frac{dr}{\sqrt{1-2M/r}}
\end{equation}
is the surface stress energy  of the shell. 
(See e.g\,, \cite{MTW}.)

\subsection{Even parity equations of motion of the matter in the shell}

From Eq.\,(\ref{firstorder}), we obtain the equations of motion of the shell,
$\delta T^{\alpha\beta}\:_{\rm shell;\beta}=0$, where $;$ denotes the covariant
derivative in the Schwarzschild spacetime. Taking Eq.\,(\ref{shellstress}) into
account leads to the restriction $S^{r\alpha}=0$ and to the
partial differential equations,
\begin{equation}\label{shellequations1}
S^{ab}\:_{|b}=0,\quad S^{r\alpha}\:_{;\alpha}=0, \quad a,b=t,\theta,\varphi
\end{equation}
for the components of the surface stress energy tensor.  Here
``$_{|}$'' is the covariant derivative with respect to the shell
3-metric in Eq.\,(\ref{shellmetric}).

Due to the spherical symmetry of the shell, we can decompose $S_{00},
S_{0i},S_{ij}; i,j=\theta,\varphi$ in scalar, vector and tensor
spherical harmonics respectively. Restricting attention to the even
parity harmonics, we write these decompositions as:
\begin{mathletters}\label{evenstress}
\begin{eqnarray}
& &
S_{00}=\sum_{l}S_{l0}^{3}(t)Y_{l0}(\theta)\\
& &
S_{0\theta}=\sum_{l}S_{l0}^{4}(t)\frac{\partial}{\partial\theta}
Y_{l0}(\theta)\\
& &
S_{ij}=\sum_{l}[S^{5}_{l0}(t)\Phi_{l0ij}(\theta)+S^{6}_{l0}(t)
\Psi_{l0ij}(\theta)]\ .
\end{eqnarray}
\end{mathletters}
Here $Y_{l0}$ is the scalar spherical harmonic and
$\Phi_{l0\varphi\varphi}/\sin^{2}\theta=\Phi_{l0\theta\theta}=Y_{l0},
\Phi_{l0\theta\varphi}=0$ and $\Psi_{l0\theta\theta}=
\partial^{2}/\partial\theta^{2}Y_{l0},\Psi_{l0\varphi\varphi}=\sin\theta
\cos\theta\partial/\partial\theta Y_{l0},\Psi_{l0\theta\varphi}=0$ are
the even parity Regge-Wheeler \cite{RW} tensor harmonics.  Azimuthal
symmetry guarantees that the azimuthal index $m$ does not enter into
any of the equations after multipole decomposition, so with no loss of
generality we consider only $m=0$ axially symmetric motions.

Upon substitution of Eq.\,(\ref{evenstress}) into
Eq.\,(\ref{shellequations1}) we get,
\begin{mathletters}\label{shellequations2}
\begin{eqnarray}
& &
\frac{R^{2}_{\rm shell}}{1-2M/R_{\rm shell}}\frac{dS^{3}_{l0}}{dt}+l(l+1)
S^{4}_{l0}=0\\
& &
\frac{R^{2}_{\rm shell}}{1-2M/R_{\rm shell}}\frac{dS^{4}_{l0}}{dt}-
S^{5}_{l0}+\left[l(l+1)-1\right]S^{6}_{l0}=0\\
& &
\frac{MR_{\rm shell}}{(1-2M/R_{\rm shell})^{2}}S^{3}_{l0}-2S^{5}_{l0}+
l(l+1)S^{6}_{l0}=0\ .
\end{eqnarray}
\end{mathletters}
As these equations show,  we only have one degree of freedom.
The choice of  the surface mass-energy
density $S^{3}_{l0}(t)$  uniquely determines
all the other components of the shell's stress
energy through Eq.\,(\ref{shellequations2}). In this work,
we make the choice
\begin{equation}\label{gaussian}
S^{3}_{l0}(t)=\epsilon\frac{e^{-a t^{2}}}{M}\ ,
\end{equation}
where $\epsilon$ is the perturbation parameter. Since all perturbation
equations will be proportional to $\epsilon$ we will omit it
henceforth.  The use of a Gaussian time dependence for the surface
density on the shell gives us a source that it localized in time and
allows us a choice of timescale for the process that drives the
stellar fluid motions.

\section{Equations governing even parity perturbations}\label{sec:even} 
A multipole decomposition of the even parity quantities in
Eq.\,(\ref{firstorder}) leads to a set of coupled partial differential
equations in the variables $t,r$ for the coefficients of the metric
perturbation $h_{\alpha\beta}$ and velocity of the fluid
star\cite{THOCAMP}\cite{Z2}.
We adopt the notation of Regge and Wheeler
\cite{RW}, Thorne and Campolattaro \cite{THOCAMP} and of Moncrief \cite{MONC}.
For simplicity,  we make the Regge-Wheeler \cite{RW} gauge choice,
which for even parity means that the only nonvanishing metric perturbations,
for a particular $l$, are
\begin{eqnarray}
& &
h_{00}=e^{\nu(r)}H_{0}^{l0}(r,t)Y_{l0}(\theta)\\
& &
h_{0r}=H_{1}^{l0}(r,t)Y_{l0}(\theta)\\
& &
h_{rr}=e^{\lambda(r)}H_{2}^{l0}(r,t)Y_{l0}(\theta)\\
& &
h_{jk}=r^{2}K^{l0}(r,t)\Phi_{l0jk}, \ \ \ \ \ \   j,k=\theta,\varphi\ ,
\end{eqnarray}
where $\Phi_{l0jk}$ is one of the even parity tensor harmonics defined
above, after Eq.\,(\ref{evenstress}). 
It is useful to divide 
the perturbed Einstein equations\,(\ref{firstorder}) 
into those governing
perturbations inside the star and those for
perturbations outside the star.

\subsection{The interior equations}
Inside the star the only nonzero stress energy is due to the perturbed fluid.
Its independent components, are
\begin{mathletters}\label{evenfluid}
\begin{eqnarray}
& &
\delta T_{0}\:^{0}_{\rm fluid}=-\delta\rho\\
& &
\delta T_{r}\:^{r}_{\rm fluid}=\delta T_{\theta}\:^{\theta}_{\rm fluid}=
\delta T_{\varphi}\:^{\varphi}_{\rm fluid}=\delta p\\
& &
\delta T_{0}\:^{a}_{\rm fluid}=(\rho+p)u_{0}\delta u^{a},\\
& &
\delta T_{a}\:^{0}_{\rm fluid}=(\rho+p)\delta u_{a}u^{0},  \ 
\ \ \ \ \  a=r,\theta,\varphi\ ,
\end{eqnarray}
\end{mathletters}
where $\delta\rho,\delta p$ are the Eulerian changes in density and
pressure, $u^{0}=e^{-\nu(r)/2}$ is the only nonzero component of the
velocity of the unperturbed star and $\delta u^{\alpha}$ is the
velocity of the perturbed fluid. It is convenient to introduce, at
this point, the quantity
\begin{equation}\label{enthalpy}
\delta h\equiv \frac{\delta p}{\rho+p}\ .
\end{equation}
For barotropic fluids, $\delta h$ is the Eulerian perturbation of the 
relativistic enthalpy and
\begin{equation}\label{reldens}
\delta \rho=\frac{(p+\rho)^{2}}{p\gamma}\delta h\ ,
\end{equation}
where $\gamma$ is the adiabatic index,
\begin{equation}\label{adiabatic}
\gamma\equiv \frac{p+\rho}{p}\;\frac{dp/dr}{d\rho/dr}\ .
\end{equation}
We decompose the stress energy components in Eqs.\,(\ref{evenfluid})
into spherical harmonics, and for a single multipole have
\begin{eqnarray}
& &
\delta h=\delta h_{l}(r,t)\,Y_{l0}(\theta)\\
& &
\delta u^{0}=\frac{1}{2}e^{-\nu(r)/2}H_{0}^{l0}(r,t)\,Y_{l0}(\theta)\\
& &
\delta u^{r}=-\frac{e^{-[\nu(r)+\lambda(r)]/2}}{r^{2}}
\,\frac{\partial}{\partial t} 
W_{l0}(r,t)\,Y_{l0}(\theta)\\
& &
\delta u^{\theta}=\frac{e^{-\nu(r)/2}}{r^{2}}\frac{\partial}{\partial t}
V_{l0}(r,t)\,\frac{\partial}{\partial\theta}\,Y_{l0}(\theta)\ .
\end{eqnarray}

With a similar decomposition of the perturbed Einstein tensor into
tensor harmonics, Eq.\,(\ref{firstorder}) leads to a set of coupled
equations for $H_{0}^{l0}, H_{1}^{l0}, K^{l0}, H_{2}^{l0}, W_{l0},
V_{l0}$ and $\delta h_{l0}$. One of the equations\cite{THOCAMP}
provides the important simplification,
\begin{equation}\label{reduction1}
H_{2}^{l0}(r,t)=H_{0}^{l0}(r,t)\ .
\end{equation}
Using this result, the other equations,  can be reduced to a coupled system of two
equations\cite{IP} in which all fluid functions have been eliminated and the
only dependent variables are the metric functions $H_{0}^{l0},
K^{l0}$. For barotropic fluids, these equations  are
\begin{eqnarray}
& &\nonumber
e^{[\lambda(r)-3\nu(r)]/2}[e^{[3\nu(r)-\lambda(r)]/2}K_{,r}^{l0}]_{,r}+
2\left(-\frac{1}{r}+\frac{1}{2}\nu_{,r}\right)K_{,r}^{l0}-e^{\lambda(r)-\nu(r)}
K_{,tt}^{l0}-(l-1)(l+2)\frac{e^{\lambda(r)}}{r^{2}}K^{l0}-H_{0,rr}^{l0}\\
& &\nonumber
+\left(\frac{2}{r}+\frac{\lambda_{,r}}{2}-\frac{5\nu_{,r}}{2}\right)H_{0,r}^{l0}
+\left\{\frac{l(l+1)}{r^{2}}e^{\lambda(r)}-\frac{2}{r^{2}}(1-r\nu_{,r})-
e^{[\lambda(r)-3\nu(r)]/2}[e^{[3\nu(r)-\lambda(r)]/2}\nu_{,r}]_{,r}\right.\\
& &\label{firstfluid}
\left.-8\pi e^{\lambda(r)}(\rho+p)\right\}H_{0}^{l0}+e^{\lambda(r)-\nu(r)}
H_{0,tt}^{l0}=0\ ,
\end{eqnarray}
and
\begin{eqnarray}
& &\nonumber
K_{,rr}^{l0}+\left[\frac{3}{r}-\frac{\lambda_{,r}}{2}+\frac{p+\rho}{p\gamma}
\left(-\frac{1}{r}+\frac{1}{2}\nu_{,r}\right)\right]K_{,r}^{l0}-
\frac{e^{\lambda(r)}}{2r^{2}}(l-1)(l+2)\left(1+\frac{p+\rho}{p\gamma}\right)
K^{l0}-e^{\lambda(r)-\nu(r)}\frac{p+\rho}{p\gamma}K_{,tt}^{l0}\\
& &\label{secondfluid}
-\left(1-\frac{p+\rho}{p\gamma}\right)\frac{H_{0,r}^{l0}}{r}-
\left[-\frac{\lambda_{,r}}{r}+\frac{1}{r^{2}}+\frac{l(l+1)}{2r^{2}}e^{\lambda(r)}
+\frac{p+\rho}{p\gamma}\left(\frac{1}{r^{2}}-\frac{l(l+1)}{2r^{2}}e^{\lambda(r)}
\right)-\frac{\nu_{,r}}{r}\frac{p+\rho}{p\gamma}\right]H_{0}^{l0}=0\ .
\end{eqnarray}
For HIF models, the adiabatic index $\gamma$ is effectively infinite  since
$\rho=$constant, and thus all the terms proportional to
$(p+\rho)/p\gamma$ in Eq.\,(\ref{secondfluid}) are zero.
\subsection{Reduction of the exterior equations to the Zerilli equation}
In the Schwarzschild spacetime outside  the star, 
it can be shown \cite{Z2}\cite{Z1}
that all the perturbation equations can be obtained from the two
first order equations
\begin{eqnarray}
& &\label{first1}
l(l+1)H_{1}^{l0}+2rH_{0,t}^{l0}-2r^{2}K_{,tr}^{l0}+\frac{6M-2r}{1-2M/r}
K_{,t}^{l0}=D_{l0}(r,t)\\
& &\label{first2}
\frac{2M}{r^{2}}H_{1}^{l0}+\left(1-\frac{2M}{r}\right)H_{1,r}^{l0}-K_{,t}^{l0}
-H_{0,t}^{l0}=B_{l0}(r,t)
\end{eqnarray}
together with the algebraic identity
\begin{eqnarray}
& &\nonumber
F\equiv \left[(l-1)(l+2)+\frac{6M}{r}\right]H_{0,t}^{l0}-
\left[(l-1)(l+2)+\frac{2M(r-3M)}{r(r-2M)}\right]K_{,t}^{l0}\\
& &\label{algebraic}
-\frac{2r^{2}}{1-2M/r}K_{,ttt}^{l0}
+2rH_{1,tt}^{l0}+M\frac{l(l+1)}{r^{2}}H_{1}^{l0}=0
\end{eqnarray}
and the shell's equations of motion, Eq.\,(\ref{shellequations2}). The
source terms in Eqs.\,(\ref{first1}), (\ref{first2}) are
\begin{eqnarray}
& &
D_{l0}=32\pi r\frac{dS^{6}_{l0}}{dt}\sqrt{1-\frac{2M}{r}}
\delta\left(r-R_{\rm shell}\right)\\
& &
B_{l0}=16\pi \left(S^{4}_{l0}-\frac{dS^{6}_{l0}}{dt}\right)\sqrt{1-\frac{2M}{r}}
\delta\left(r-R_{\rm shell}\right)\ .
\end{eqnarray}

Equations (\ref{first1}), (\ref{first2}) and (\ref{algebraic}),
in turn, can be combined into the single wave equation,
\begin{equation}\label{zereq}
\frac{\partial^{2}Z_{l0}}{\partial t^{2}}-\frac{\partial^{2}Z_{l0}}
{\partial r^{*2}}+V_{l}(r)Z_{l0}={\cal S}_{l0}(r,t)
\end{equation}
for the Zerilli \cite{Z1} function,
\begin{equation}\label{zerfunction}
Z_{l0}(r,t)\equiv\frac{r(r-2M)}{(r\lambda+3M)(\lambda+1)}[H_{0}^{l0}-rK^{l0}_{,r}]
+\frac{r}{\lambda+1}K^{l0}\ .
\end{equation}
In Eq.\,(\ref{zereq}), $r^{*}$ is the usual {\it tortoise} coordinate,
\begin{equation}
r^{*}=r+2M\log[r/2M-1]+{\rm constant}\ ,
\end{equation}
the constant $\Lambda$  is
\begin{equation}
\Lambda\equiv\frac{(l-1)(l+2)}{2},
\end{equation}
and the potential $V_{l}$ is
\begin{equation}\label{outpotential}
V_{l}(r)=\frac{2\Lambda^{2}(\Lambda+1)r^{3}+6\Lambda^{2}Mr^{2}+18\Lambda M^{2}
r+18M^{3}}{r^{3}(\Lambda r+3M)^{2}}\left(1-\frac{2M}{r}\right)\ .
\end{equation}
The source term in Eq.\,(\ref{zereq}) is
\begin{eqnarray}
& &\nonumber
{\cal S}_{l0}(r,t)=-\frac{C_{l0}(r,t)f(r)}{h(r)}\delta[r-R_{\rm shell}]
+\frac{\partial}{\partial r^{*}}\left[\frac{C_{l0}(r,t)\delta[r-R_{\rm shell}]}{
h(r)}\right]\\
& &\label{source}
+\frac{16\pi}{R_{\rm shell}}\left(1-\frac{2M}{R_{\rm shell}}\right)^{3/2}
S^{6}_{l0}(t)\delta\left(r-R_{\rm shell}\right)\ ,
\end{eqnarray}
where
\begin{eqnarray}
&&C_{l0}(r,t)=-\frac{16\pi R^{2}_{\rm shell}}{l(l+1)
\sqrt{1-2M/R_{\rm shell}}}S^{3}_{l0}(t)\\
& &
f(r)=\frac{6M^{2}+3\Lambda Mr+r^{2}\Lambda(\Lambda+1)}{r^{2}(3M+\Lambda r)}\\
& &
h(r)=\frac{3M+\Lambda r}{r-2M}\ .
\end{eqnarray}
Together with appropriate boundary conditions, the set of equations
(\ref{firstfluid}), (\ref{secondfluid}) and (\ref{zereq}) 
govern the even parity perturbations of the star due to the shell.

\subsection{Boundary conditions}
\subsubsection{Regularity at the center of the star}
The system of equations (\ref{firstfluid}-\ref{secondfluid}) admits
two linearly independent regular solutions. Near the center they admit
the series expansion\cite{IP},
\begin{mathletters}\label{atcenter}
\begin{eqnarray}
& &
K^{l0}(r,t)=k_{0}(t)r^{l}+k_{2}(t)r^{l+2}+O(r^{l+4})\\
& &
H_{0}^{l0}(r,t)=k_{0}(t)r^{l}+h_{2}(t)r^{l+2}+O(r^{l+4})\ .
\end{eqnarray}
\end{mathletters}
The $r$-independent ``constants,'' $k_{0}$, 
$k_{2}$, $h_{2}$ are {\em a priori}
arbitrary. When 
Eq.\,(\ref{secondfluid}) is expanded about $r=0$ and the above
expansions are used, we find
\begin{equation}\label{constraint1}
-\left[\frac{3}{2}l+3+\frac{l^{2}}{2}\right]h_{2}+\left[\frac{l^{2}}{2}+
\frac{11}{2}l+9\right]k_{2}-\frac{8\pi}{3}\rho_{0}\left[2l+l^{2}-3\right]k_{0}=0
\end{equation}
where $\rho_{0}, p_{0},\gamma_{0}$ are the values of the density and
pressure at the center of the star and we used the fact that
$\gamma=\infty$ for a HIF. Equation 
(\ref{constraint1})
allows us to eliminate $k_{0}$,
so that the only remaining unknown coefficients are $k_{2}$ and
$h_{2}$. These  turn out to be fixed, up to an overall
scaling, by conditions at the stellar surface.

\subsubsection{Vanishing of the pressure at the stellar surface}
The pressure must vanish at the perturbed surface of the star by
definition of that boundary. This amounts to the vanishing of the
Lagrangian perturbation of the pressure at the stellar surface $r=R$,
\begin{equation}\label{pressure1}
\delta p(R)-\frac{e^{-\lambda(R)/2}}{R^{2}}W(R,t)p_{,r}(R)=0\ .
\end{equation}
When the density $\rho$ vanishes at $r=R$, the second term in
Eq.\,(\ref{pressure1}) is zero and this condition is the same as the
vanishing of the Eulerian perturbation of the pressure. But this is
not so when $\rho(R)\neq 0$ as is the case for a HIF.

After differentiating Eq.\,(\ref{pressure1}) twice with respect to
time and doing a decomposition in spherical harmonics, we can rewrite
it purely in terms of the metric functions $K^{l0}, H_{0}^{l0}$ and
its derivatives\cite{IP}. The resulting expression, when $\rho(R)\neq
0$, is
\begin{eqnarray}
& &\nonumber
\left\{-\frac{1}{r}K_{,ttr}^{l0}-\frac{\nu_{,r}}{4}\frac{l(l+1)}{r^{2}}e^{\nu(r)}
K_{,r}^{l0}-\frac{1}{2}\frac{(l-1)(l+2)}{r^{2}}e^{\lambda(r)}K_{,tt}^{l0}-
e^{\lambda(r)-\nu(r)}K_{,tttt}^{l0}-\frac{1}{2}\frac{\nu_{,r}}{r}
\left(1-r\frac{\nu_{,r}}{2}\right)K_{,tt}^{l0}\right.\\
& &\label{pressure2}
\left.+\frac{1}{r}H_{0,rtt}^{l0}+\frac{l(l+1)}{4r^{2}}
e^{\nu(r)}\nu_{,r}H_{0,r}^{l0}
+\left[3\frac{\nu_{,r}}{2r}-\frac{1}{r^{2}}+
\frac{l(l+1)}{2r^{2}}e^{\lambda(r)}\right]H_{0,tt}^{l0}+(\nu_{,r})^{2}
e^{\nu(r)}\frac{l(l+1)}{4r^{2}}H_{0}^{l0}\right\}_{r=R}=0\ .
\end{eqnarray}

\subsubsection{The relation between the interior and the exterior metric
functions}
If the density of the star is not zero at $r=R$, the first radial
derivative of $K^{l0}, H_{0}^{l0}$ is {\it not} continuous at $r=R$,
although the functions themselves are. Denoting by a superscript
``$+$'' the exterior metric functions and by a superscript ``$-$'' the
interior ones, we have\cite{IP}, for a particular multipole $l$, that
\begin{mathletters}\label{matching}
\begin{eqnarray}
& &
K^{+}(R,t)=K^{-}(R,t)\\
& &
H_{0}^{+}(R,t)=H_{0}^{-}(R,t)\\
& &\nonumber
K_{,rtt}^{+}(R,t)=K_{,rtt}^{-}(R,t)-\frac{1}{2R^{2}}\left\{
l(l+1)\left(1-\frac{2M}{r}\right)K_{,r}^{-}+2r\left[\frac{r-3M}{r-2M}
K_{,tt}^{-}+rK_{,rtt}^{-}\right]\right.\\
& &
\left.-l(l+1)\left(1-\frac{2M}{r}\right)H_{0,r}^{-}-2rH_{0,tt}^{-}-2M
\frac{l(l+1)}{r^{2}}H_{0}\right\}_{r=R}\ .
\end{eqnarray}
\end{mathletters}
If 
$\rho(R)=0$ then both metric functions and their first radial derivatives 
would be
continuous at $r=R$.

\subsection{Fourier transform of the perturbation equations}
The simplest way to solve the above partial differential equations in
$r,t$ is to write all time dependent quantities as Fourier integrals,
reducing the problem to that of ordinary differential equations in
$r$. Thus we write $Z_{l0}(r,t)$ as,
\begin{equation}
Z_{l0}(r,t)=\frac{1}{2\pi}\int_{-\infty}^{\infty}e^{-i\omega t}\tilde{Z}_{l0}
(r,\omega)d\omega
\end{equation}
transforming Eq.\,(\ref{zereq}) into a second order equation for 
$\tilde{Z}_{l0}(r,\omega)$,
\begin{equation}\label{zereqfour}
\frac{d^{2}\tilde{Z}_{l0}}{dr^{*2}}+[\omega^{2}-V_{l}(r)]\tilde{Z}_{l0}=-
\tilde{{\cal S}}_{l0}(r,\omega)
\end{equation}
where
\begin{equation}\label{sourcefour}
\tilde{{\cal S}}_{l0}(r,\omega)=\int_{-\infty}^{\infty}
{\cal S}_{l0}(r,t)e^{i\omega t} dt
\end{equation}
is the Fourier transform of the source term. We can proceed similarly
with Eqs.\,(\ref{firstfluid}) and (\ref{secondfluid}). The resulting
equations for $\tilde{H}_{0}^{l0}(r,\omega), \tilde{K}^{l0}(r,\omega)$
can be obtained directly from Eqs.\,(\ref{firstfluid}) and
(\ref{secondfluid}) by substituting $-i\omega$ for $\partial_{t}$, and
by replacing functions of $r,t$ by their Fourier transforms.

At infinity, we impose the boundary condition that the wave be purely
outgoing
\begin{equation}\label{zeratinfty}
Z_{l0}(r\rightarrow\infty,u\equiv
t-r^{*})=\frac{1}{2\pi}\int_{-\infty}^{\infty}
A_{l0}(\omega)e^{-i\omega u}\;d\omega\ .
\end{equation}
The even parity gravitational energy, radiated in a
single multipole component, is then given\cite{CPM} by
\begin{equation}\label{spectrum}
\frac{dE_{l}}{d\omega}=\frac{1}{64\pi^{2}}\frac{(l+2)!}{(l-2)!}\omega^{2}
|A_{l0}(\omega)|^{2}\ .
\end{equation}
From the wave function $Z_{l0}(r,t)$ we can construct the multipoles
of the metric perturbations. The transverse traceless (TT) part of the
metric perturbation is of particular interest, since it completely
characterizes the radiation at infinity. Extracting the TT part is
easiest if we realize that the Zerilli function $Z_{l0}(r,t)$ is
numerically equal to Moncrief's \cite{MONC} gauge invariant
wavefunction. Since the Moncrief invariant can be evaluated in any
gauge, we choose the gauge to be asymptotically flat so that all
multipole perturbations fall off faster than $1/r$ except the TT
components. From this procedure we find that the TT perturbations are
related to the Zerilli function by
\begin{equation}
h_{jk}^{TT}=\frac{1}{r}\sum_{l=2}Z_{l0}(r,t)\sqrt{\frac{(l+2)!}{2(l-2)!}}
\;T^{E2,l0}_{jk}(\theta), \ \ \ \ \ \
j,k=\theta,\varphi\ ,
\end{equation}
where
\begin{equation}
T^{E2,l0}_{jk}(\theta)=\sqrt{2\frac{(l-2)!}{(l+2)!}}
\,\left[\Psi_{l0jk}+\frac{l(l+1)}{2}
\Phi_{l0jk}\right]
\end{equation}
is the orthonormal even parity TT tensor harmonic and
$\Psi_{l0},\Phi_{l0}$ are the tensor harmonics introduced in
Eq.\,(\ref{evenstress}).
\section{Computational implementation}\label{sec:compimp} 
\subsection{Solution for $A_{l0}(\omega)$}

If the background spacetime is due to a star, a solution of
Eq.\,(\ref{zereqfour}) must be found that corresponds to outgoing
waves at infinity and that matches the regular solution of
Eq.\,(\ref{firstfluid}), (\ref{secondfluid}) at the unperturbed
surface of the star according to the junction conditions of
Eq.\,(\ref{matching}). The Green function solution is found in the
usual way. (See e.g.\,, Ref.\,\cite{ZP}.) We define
$y^{out}_{l}(r,\omega)$ as the homogeneous solution of
Eq.\,(\ref{zereqfour}) with the asymptotic form
\begin{equation}
y^{out}_{l}(r,\omega)\rightarrow e^{i\omega r^{*}}\quad\quad r^{*}\rightarrow
\infty\ .
\end{equation}
For our second independent solution of Eq.\,(\ref{zereqfour}) we start
by finding, in the stellar interior, a solution of
Eqs.\,(\ref{firstfluid}), (\ref{secondfluid}) satisfying the condition
Eq.\,(\ref{atcenter}) at the stellar center.  Our second solution
$y^{reg}_{l}(r,\omega)$ is taken to be the homogeneous solution of
Eq.\,(\ref{zereqfour}) that joins to the interior solution through the
matching conditons of Eq.\,(\ref{matching}).

We then define the Wronskian of these two homogeneous solutions, an $r$
independent quantity, to be
\begin{equation}\label{wronskian}
W_{l}(\omega)=y^{reg}_{l}\frac{dy^{out}_{l}}{dr^{*}}-y^{out}_{l}
\frac{dy^{reg}_{l}}{dr^{*}}\ .
\end{equation}
With the above definitions, and from the Green function solution, we
obtain (see, e.g.\,,Ref.\,\cite{ZP}) the Fourier amplitude $A_{l0}$
defined in Eq.\,(\ref{zeratinfty})
\begin{equation}\label{ALzero} 
A_{l0}(\omega)=-\frac{1}{W_{l}(\omega)}\int_{r<R_{\rm shell}}^{\infty}
\tilde{{\cal S}}_{l0}(r,\omega)\frac{y^{reg}_{l}(r,\omega)}{1-2M/r}dr\ .
\end{equation}
Combining Eq.\,(\ref{shellequations2}), (\ref{gaussian}), (\ref{source}) and
(\ref{sourcefour}) we get explicitly,
\begin{eqnarray}
& &\nonumber
A_{l0}(\omega)=-\frac{16\pi}{W_{l}(\omega)\sqrt{1-2M/R_{\rm shell}}}
\sqrt{\frac{\pi}{a}}e^{-\omega^{2}/(4a)}\left\{\frac{R^{2}_{\rm shell}}
{Mh(r)l(l+1)}\frac{dy^{reg}}{dr}(R_{\rm shell},\omega)\right.\\
& &\label{explicit}
\left.+\frac{y^{reg}_{l}(R_{\rm shell},\omega)}{1-2M/R_{\rm shell}}\left
[\frac{1}{(l-1)(l+2)}\left(1-\frac{2R^{3}_{\rm shell}\omega^{2}}{Ml(l+1)}
\right)+
\frac{R^{2}_{\rm shell}f(r)}{Mh(r)l(l+1)}\right]\right\}\ .
\end{eqnarray}
What is required for a solution, then, is a numerical determination of
$y^{reg}_{l}(R_{\rm shell},\omega)$ and its radial derivative.

\subsection{Numerical method to find $y^{reg}$ for a HIF}
The numerical problem of finding $y^{reg}_{l}(R_{\rm shell},\omega)$
and its derivative for a HIF, can be divided in two parts:
Integration of Eqs.\,(\ref{firstfluid}) and (\ref{secondfluid}) from
the center of the star to $r=R$ and integration of
Eq.\,(\ref{zereqfour}) from $r=R$ to  $r=R_{\rm shell}$, so that we
can evaluate $y^{reg}_{l}, dy^{reg}_{l}/dr$ at this radius.

To find $K^{-}(R,\omega), H_{0}^{-}(R,\omega)$ and
$K^{-}_{,r}(R,\omega)$, for a HIF we first
find an ``A solution'' by starting from the center with with $k_{0}=1,
k_{2}=0$ and $h_{2}$ obtained from Eq.\,(\ref{constraint1}); we then 
integrate Eqs.\,(\ref{firstfluid}), (\ref{secondfluid}) out to 
$r=R$, to find the ``A solution'' there. The procedure for the ``B solution'' 
at $r=R$ is the same except we start with the central conditions
$k_{0}=0, k_{2}=1$. The general solution for $H_{0}^{-}$ and $K^{-}$
can be written 
\begin{equation}\label{generalsol}
H_{0}^{-}=\alpha\;H_{0}^{A-}+\beta\;H_{0}^{B-}\quad\quad 
K^{-}=\alpha\;K^{A-}+\beta\;K^{B-}\ .
\end{equation}
where $\alpha$ and $\beta$ are arbitrary constants.  The overall scale
of $y^{reg}_{l}(r,\omega)$ is arbitrary. Since $y^{reg}_{l}(r,\omega)$
occurs both in the integrand in Eq.\,(\ref{ALzero}) and in the
Wronskian in the denominator, the scale cancels out so only the ratio
$\alpha/\beta$ is of importance. This ratio can be found by
substituting the expression in Eq.\,(\ref{generalsol}) and its 
derivatives in the vanishing of the Lagrangian pressure condition
Eq.\,(\ref{pressure2}).

We can next compute $K^{+}(R,\omega), H^{+}(R,\omega)$ and
$K^{+}_{,r}(R,\omega)$ using Eq.\,(\ref{matching}). The final step is
to use Eq.\,(\ref{zerfunction}) to find the starting values 
for integrating the Zerilli equation,
\begin{equation}
y^{reg}_{l}(R,\omega)=\frac{R(R-2M)}{(R\Lambda+3M)(\Lambda+1)}[H_{0}^{l0+}
-rK^{l0+}_{,r}]_{r=R}+
\frac{r}{\Lambda+1}K^{l0+}(R,\omega)
\end{equation}
\begin{equation}
\frac{dy^{reg}_{l}(R,\omega)}{dr}=\frac{3M}{(R\Lambda+3M)(\Lambda+1)}
K^{+l0}(R,\omega)+\frac{R[6M^{2}+3M\Lambda R+R^{2}\Lambda(\Lambda+1)]}
{(R\Lambda+3M)^{2}(\Lambda+1)}\left[K_{,r}^{l0+}-\frac{H_{0}^{l0+}}{R}
\right]_{r=R_{\rm shell}}\ ,
\end{equation}
and to integrate the Zerilli equation Eq.\,(\ref{zereq}) out to
$r=R_{\rm shell}$ in order to find $y^{reg}_{l}(R_{\rm shell},\omega)$
and $dy^{reg}_{l}/dr(R_{\rm shell},\omega)$.

\subsection{Radiation due only to  the shell}\label{shellonly} 

The waveform $Z_{l0}$, its Fourier transform $A_{l0}(\omega)$, and the
energy computed in Eq.\,(\ref{spectrum}) refer, of course, to
radiation from the star and the shell. As might be expected the
radiation that can be attributed directly to the shell is much larger
than the radiation from the perturbed fluid motions in the neutron
star.  In order to have the clearest comparison of Newtonian and
relativistic predictions of radiation from the neutron star, it is
useful to remove the contribution due to the shell.

For the Newtonian computation presented in the appendix, this presents
no problem. As is evident in Eq.\,(\ref{quaddens}), the density
perturbations due to the shell and due to the star are distinct, and
it is evident in Eq.\,(\ref{quadmom}) that their contribution to the
quadrupole moment are distinct.  This clear distinction does not exist
in the relativistic calculation. The equations of Sec.\,\ref{sec:even}
contain information about the shell entangled with information about
the star. To get an approximate idea of what radiation can be ascribed
to the star, and not to the shell, one can compute the waveform due
(in a sense) to the shell itself, and subtract this waveform from the
total star+shell waveform.  This subtraction, however, is somewhat
subtle.

In particular, it is not useful to consider the shell in a flat
spacetime background. If we consider a shell with the surface stress
energy of Eqs.\,(\ref{shellequations2}), (\ref{gaussian}) radiating in
a flat background, and the same (in some sense) shell, and surface
stresses, radiating in another fixed background, there will be a large
difference simply due to the different radial null geodesics, the
spacetime lines on which perturbations propagate.  If we are to have
the same shell-only radiation as that due to the shell in the
shell+star problem, we must have the two signals propagate on the same
spacetime background. To accomplish this, we describe the waves at
radius $r>R$ for the shell-only problem with the same Zerilli equation
as we use in the shell+star problem.  Equation (\ref{zereq}) is then
part of both problems for $r>R$. In the shell+star problem the
perturbations in the interior are, of course, treated with the
equations of Sec.\,\ref{sec:even}.  For the shell-only problem, we
need instead something like a Zerilli equation suitable to the fixed
spacetime of the stellar interior. Such an equation is provided by the
equation for the propagation of massless scalar perturbations in the
neutron star background. These waves do not excite oscillations either
of the spacetime or of the neutron star fluid. They are described only
by an equation identical in form to Eq.\,(\ref{zereq}), but with no
source, and with the Zerilli potential of Eq.\,(\ref{outpotential})
replaced by
\begin{equation}\label{scalarpot} 
V^{\rm scalar}_{l}=e^{\nu}\left[\frac{l(l+1)}{r^{2}}+
\frac{\nu_{,r}-\lambda_{,r}}{2r}e^{-\lambda}\right]\ .
\end{equation}
(The idea of freezing the fluid perturbations to study the $w$ modes
described in \cite{AKS2} is somewhat similar in spirit to our method
but is different in practice.)  With this mixed mathematical
description we find the regular solution of the Zerilli equation and
then compute, using Eq.\,(\ref{explicit}), the Fourier transform
$A_{l0}^{\rm shell}$ and, with Eq.\,(\ref{zeratinfty}), a wave
function $Z_{l0}^{\rm shell}$. By subtracting these, respectively from
the $A_{l0}$ and $Z_{l0}$ for the shell+star, we arrive at results
$A_{l0}^{\rm star}$, $Z_{l0}^{\rm star}$ meant to describe radiation
only due to the oscillations of the neutron star fluid. From the
square of $A_{l0}^{\rm star}$, we can compute an energy spectrum and a
total radiation energy attributed to the motion of the stellar fluid.

Our method of finding the radiation from the shell alone is, of
course, only an approximation. Einstein's equations couple
oscillations of the fluid and oscillations of the spacetime, so that
there can be no completely meaningful way of separating the two. One
sign of this is that the potential Eq.\,(\ref{scalarpot}) is not
unique. The potential does not influence the radial null geodesics,
and any potential with the same general behavior at $r\rightarrow0$,
is equally good. In situations for which the details of the potential
are important, our method of computing shell-only radiation is not
justified.  But such details are not relevant for must of our
models. This can be seen in the reasonable success of our method in
removing appearances of shell radiation in the star-only results
presented below.

An additional important point to understand about our shell
subtraction method is that in principle it should subtract the $w$
modes. Since the $w$ modes are due to the spacetime background, not to
the fluid motion, the shell-only radiation should have $w$ modes, and
the star-only waves that result from subtraction should have only
waves due to fluid motion. This will be discussed further in
connection with the numerical results presented below.

\section{Numerical results}\label{sec:numres} 

\begin{figure}
\hspace*{.1\textwidth}\epsfxsize=0.35\textwidth
\epsfbox{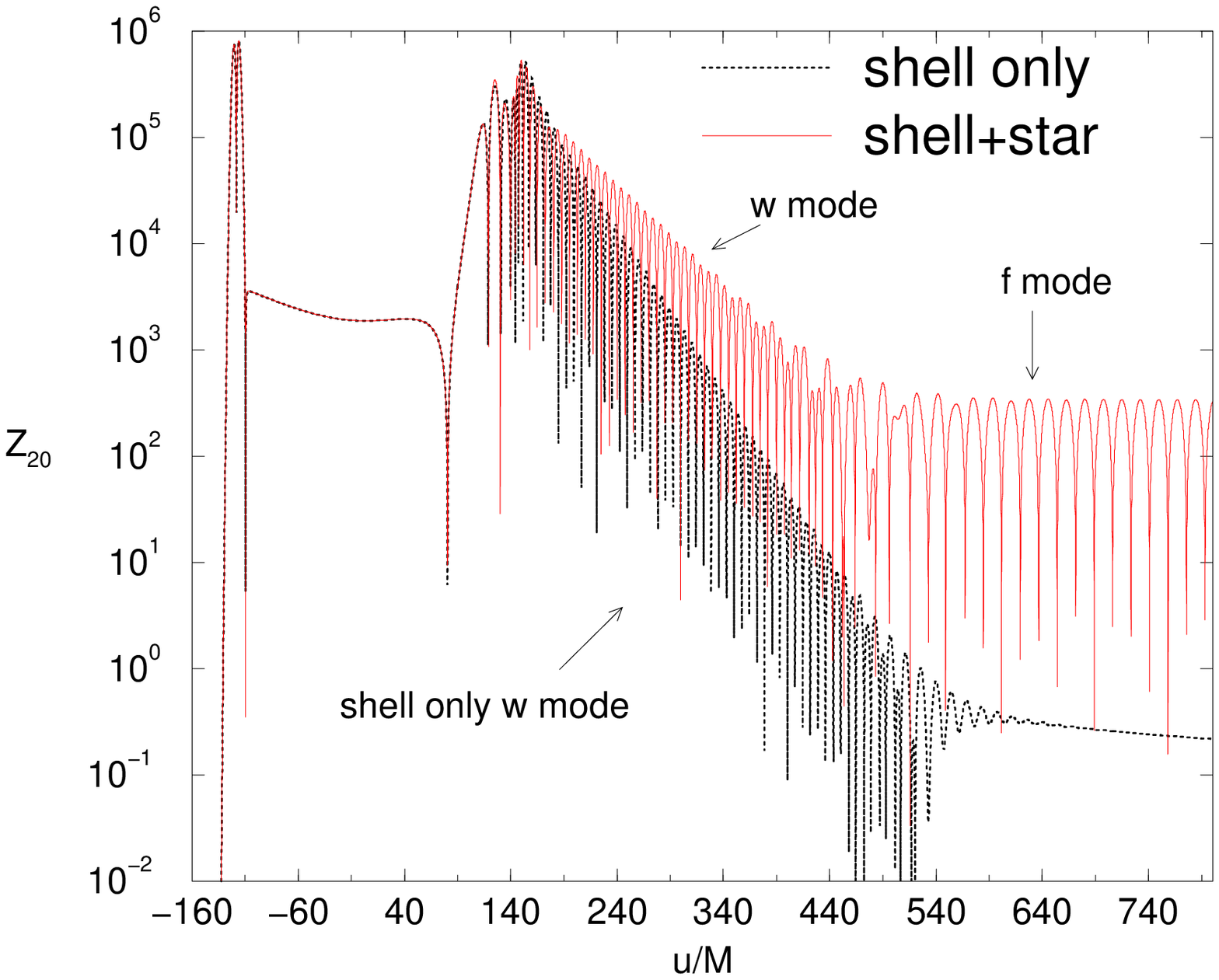}\hspace*{.1\textwidth} \epsfxsize=0.35\textwidth
\epsfbox{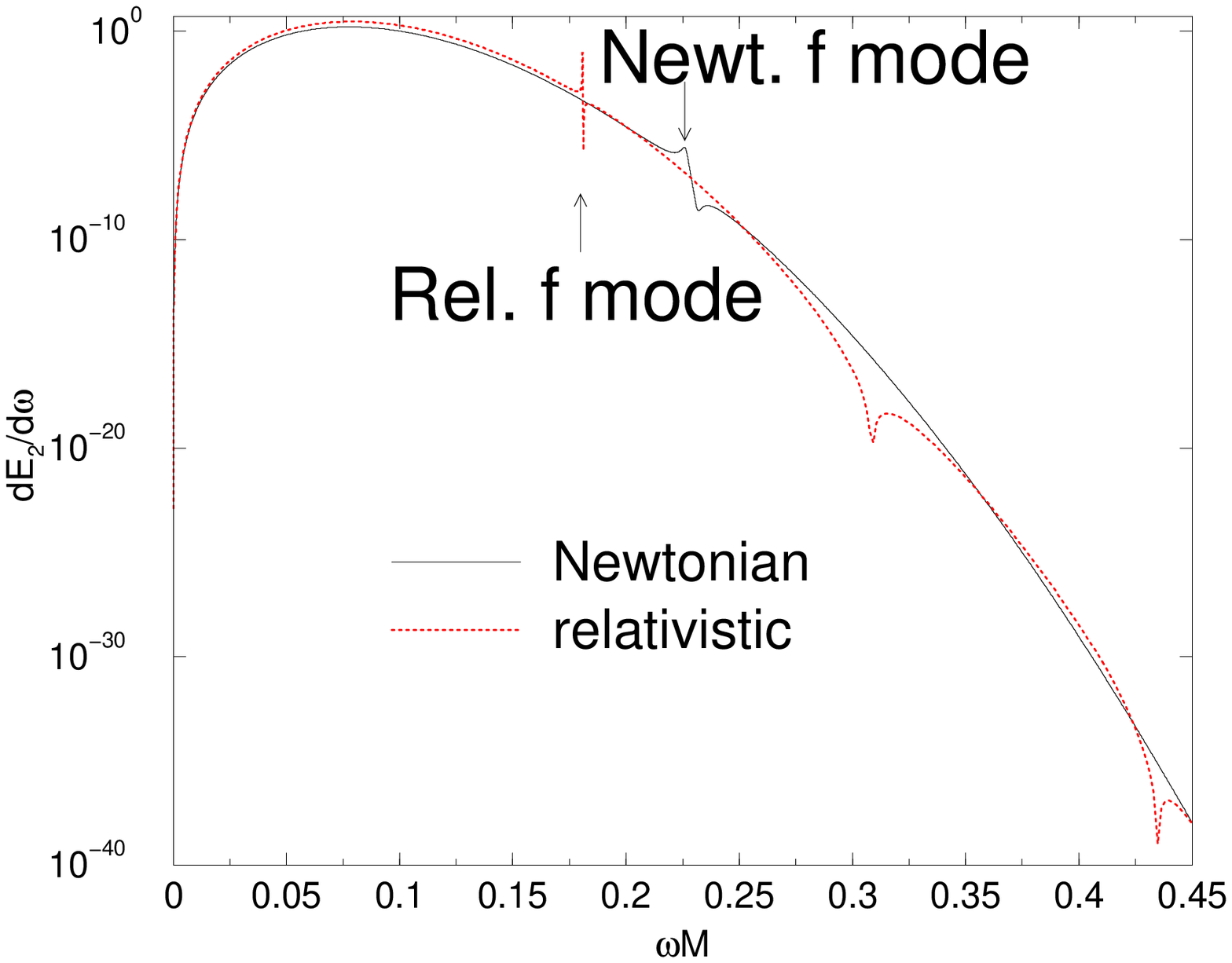} 
\caption{\label{HSR=2.5M} 
A waveform and spectrum for an extremely compact HIF model with
$R=2.5M$.  For the waveform in (a) the shell radius was taken to be
large, $R_{\rm shell}=110M$, and the time scale for the shell stress
energy small, $a=0.1M^{-2}$, in order to show a clear separation of
the initial shell radiation and the radition due to the stellar fluid.
Waveforms are shown for the relativistic star+shell, and for the
relativistic shell-only model discussed in the text.  For the spectra
in (b) the shell radius is small, $R_{\rm shell}=5M$, and the shell
time scale is slow, $a=0.001M^{-2}$, so that the model is
approximately a slow motion source. The spectrum of the relativistic
star+shell model is compared with the Newtonian shell+star model. The
agreement is good, but is dominated by the shell radiation.  For the
$R=2.5M$ model the relativistic $f$ mode is $(0.18+i0.000037)M^{-1}$,
and the Newtonian $f$ mode is $(0.226+i0.00131)M^{-1}$; the least
damped $w$ mode is $(0.42+i0.022)M^{-1}$, while the least damped $w$
mode for the shell-only model is $(0.44+i0.029)M^{-1}$.
}
\end{figure}

We focus first on a very compact HIF stellar model of radius $R=2.5M$.
By starting with an extreme, though astrophysically implausible, model
we will be able to see relativistic features that will be absent in
less compact, and more plausible, models.  In Fig.~\ref{HSR=2.5M}\,(a)
we present the waveform $Z_{20}$, the Zerilli function for
quadrupole ($l=2$) perturbations in the case that the Gaussian
parameter in Eq.\,(\ref{gaussian}) is $a=0.1M^{-2}$. The shell was
placed at a large distance from the star, $R_{\rm shell}=110M$, to
have the waveform clearly display the time profile of events. The
first burst, at around $u= t-r^*=-160M$ is radiation coming directly
from the stress energy of the shell. Later, at around $u= t-r^*\sim100M$,
a burst arrives representing the ingoing radiation from the shell that
has ``reflected'' off $r=0$. At around the same time, radiation from
oscillations of the fluid and central region of the spacetime arrive.

Two curves are shown in Fig.~\ref{HSR=2.5M}\,(a). One is the waveform
for the relativistic star+shell. The second curve is that for the
shell itself, computed by the method described in
Sec.\,\ref{shellonly}. The two curves are nearly identical up to
around $u=160M$, confirming that the early radiation is that due to
the shell. From $u=160M$ to around $400M$ or more, there are damped
oscillations of the dominant (least damped) $w$ mode.  The $w$ modes
depend on the details of wave propagation in the innermost strong
field regions, and are not the same for the star+shell problem and for
the somewhat {\em ad hoc} spacetime we constructed for the shell-only
problem. The $w$ mode frequency for the constant density $R=2.5M$
model is $\omega =(0.42+i0.0217)M^{-1}$, while that for the spacetime
of Sec.\,\ref{shellonly}, $\omega =(0.438+i0.029)M^{-1}$, is more
rapidly damped. We would, of course, find a different $w$ mode
frequency if in the shell-only problem we used a potential other than
that in Eq.\,(\ref{scalarpot}).

It is clear that there is no point in subtracting the shell-only
radiation from the star+shell. The difference waveform would contain
an artifact corresponding to the difference of the $w$ modes, and
hence would be dominated by features completely irrelevant to fluid
motions of the neutron star.  Figure~\ref{HSR=2.5M}\,(a) helps to
demonstrate that subtraction is pointless when $w$ mode radiation is
of importance in the waveform. It is only the fact that $w$ mode
radiation is a minor feature for realistically compact stars that
makes subtraction useful.

In addition to the difference in frequency of the $w$ modes there is
another, more important, difference between the two curves in
Figure~\ref{HSR=2.5M}\,(a). The star+shell curve shows an oscillation
with imperceptible damping at a frequency less than half that of the
$w$ mode. This is the $f$ mode of the fluid of the neutron star. This mode
is missing, as it should be, from the shell-only computation.

Both curves in Fig.~\ref{HSR=2.5M}(a) show relativistic results. A
comparison between the Newtonian and relativistic star+shell results
would not be of much use. The slow motion condition underlying the
Newtonian approximation would be strongly violated since the radius of
the shell $R_{\rm shell}=110M$ is much larger than the characteristic
time scale of the shell stress-energy oscillations ($\Delta t\sim
a^{-1/2}\sim$ several $M$) or of the modes of oscillation of the
spacetime or of the stellar fluid.  The Newtonian computation, based
on the slow motion approximation (time scale $\ll$ light travel time
across source) would be completely inappropriate.
In order to construct a more justifiable Newtonian comparison for a
$R=2.5M$ HIF model, we consider in Fig.~\ref{HSR=2.5M}(b), a shell
with radius $R_{\rm shell}=5M$ and with Gaussian parameter
$a=0.001M^{-2}$ (and hence timescale $\sim 30M$). The waveform in this
case is not of primary interest, since the radiation from the shell
and from the stellar fluid will not be clearly distinguishable. The
energy spectrum, however, gives the answer to the most important
questions. It shows, for example, that there is negligible difference
between the broad spectra of the Newtonian and the relativistic
results. But the broad spectrum is due to the shell.  Of more
astrophysical interest is the $f$ mode excitation. The relativistic
computation shows a lower frequency $f$ mode containing almost an
order of magnitude more radiation energy than the $f$ mode peak of the
Newtonian computation.  This conclusion, of course, applies to a model
that is too compact to be astrophysically relevant.

The feature in the relativistic spectrum at $\omega\sim0.3M^{-1}$ is
due to the repetition of the shell radiation with a time delay of
$\Delta t\sim 10M$. This produces a modulation of the form
$\cos^{2}(\omega\Delta t/2)$, and hence a dramatic decrease at
$\omega\sim\pi/\Delta t\sim 0.3$. The location of such
features is dependent on $R_{\rm shell}$ and, of course, is unrelated
to the physics of the neutron star. It is worthwhile noting that these
features are absent in the Newtonian spectrum. Since that spectrum is
based on the slow motion approximation the whole star+shell source 
is treated as if it radiates in phase, and there can be no repetition 
of the shell radiation.

\begin{figure}
\hspace*{.1\textwidth}\epsfxsize=0.35\textwidth
\epsfbox{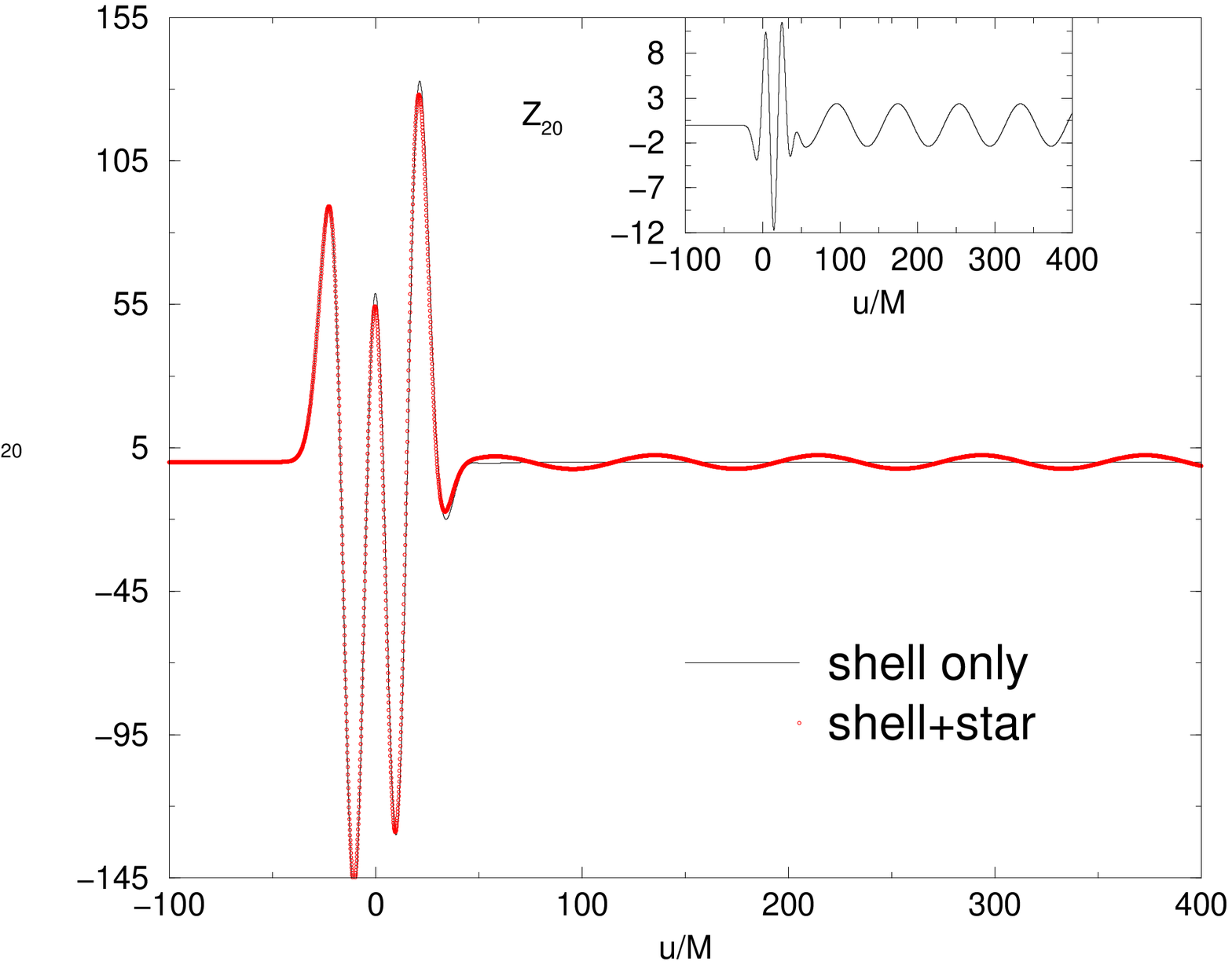}\hspace*{.1\textwidth} \epsfxsize=0.35\textwidth
\epsfbox{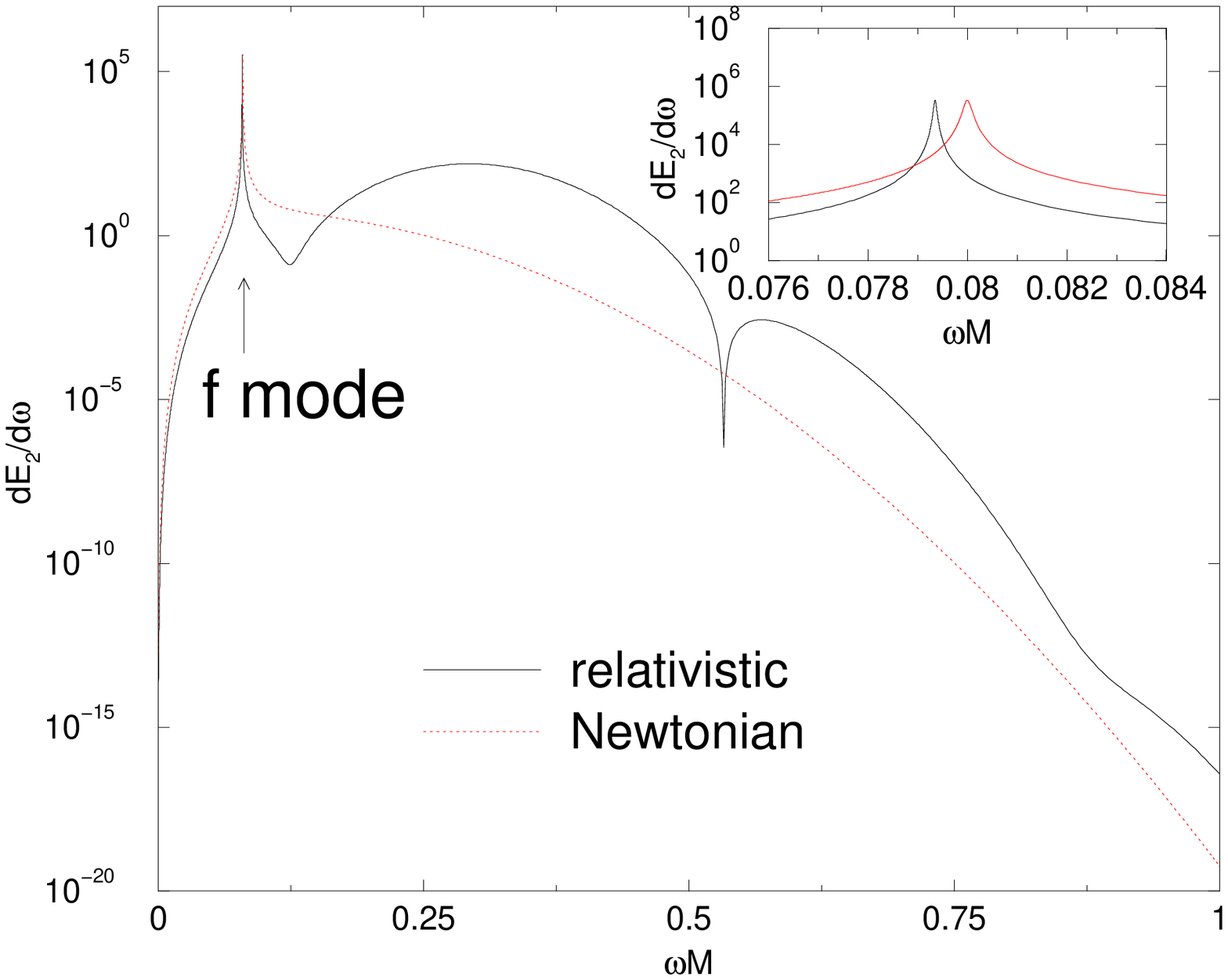} 
\caption{\label{HSR=5M} 
Results for a HIF model with $R=5M$, $R_{\rm shell}=10M$, and Gaussian
parameter is $a=0.01M^{-2}$. In (a) the radiation waveform from the
relativistic shell+star is compared to that from the relativistic
shell-only to demonstrate the value of subtraction. In (b) the
spectrum for the star-only (i.e.\,, subtracted) relativistic computation
is compared with the Newtonian star-only spectrum.
The $f$ mode frequency for the $R=5M$ model is 
 $(0.08+i0.000033)M^{-1}$ in relativity
and $(0.08+i0.000082)M^{-1}$ in the Newtonian computation.
The least damped star+shell $w$ mode frequency is
$(0.50+i0.31)M^{-1}$.
}
\end{figure}

In Fig.~\ref{HSR=5M} we show results for a HIF star of radius $R=5M$,
a typical radius of a neutron star. The shell is located at $R_{\rm
shell}=10M$ and the Gaussian parameter for the time profile of the
shell stress energy is $a=0.01M^{-2}$.  In Fig.~\ref{HSR=5M}(a) the
dotted curve gives the full Zerilli quadrupole waveform of the
star+shell computation, and shows why subtraction of shell radiation
is useful: the waveform is completely dominated by the shell
radiation, and the much smaller $f$ mode is barely visible. The solid
curve is the waveform from the shell-only computation. It is clear in
Fig.~\ref{HSR=5M}(a) that the two waveforms are nearly identical at
early times, and at late times are different in that the star+shell
result has $f$ mode oscillations, while the shell-only waveform, of
course, does not. This suggests that subtraction will be very
effective in isolating the radiation due to the stellar fluid, and
this turns out to be true. The star-only curve (the result of
subtracting the shell-only from the star+shell) is given in the inset
to Fig.~\ref{HSR=5M}(a) and shows the $f$ mode oscillation as a
dominant feature.

Figure~\ref{HSR=5M}(b) shows the energy spectrum for the subtracted
(i.e.\,, star-only) case shown in the inset of Fig.~\ref{HSR=5M}(a).
Both the relativistic and Newtonian spectra are given. The two spectra
are roughly similar in general appearance, except for the strong
feature in the relativistic spectrum at $\omega\sim0.5M^{-1}$.
Of particular interest in Fig.~\ref{HSR=5M}(b) is the excitation
of the $f$ mode (shown in detail in the inset). Perhaps the most
important feature of our model is that we can compare the excitation
given by a relativistic and a Newtonian computation. 

\begin{figure}
\hspace*{.1\textwidth}\epsfxsize=0.35\textwidth
\epsfbox{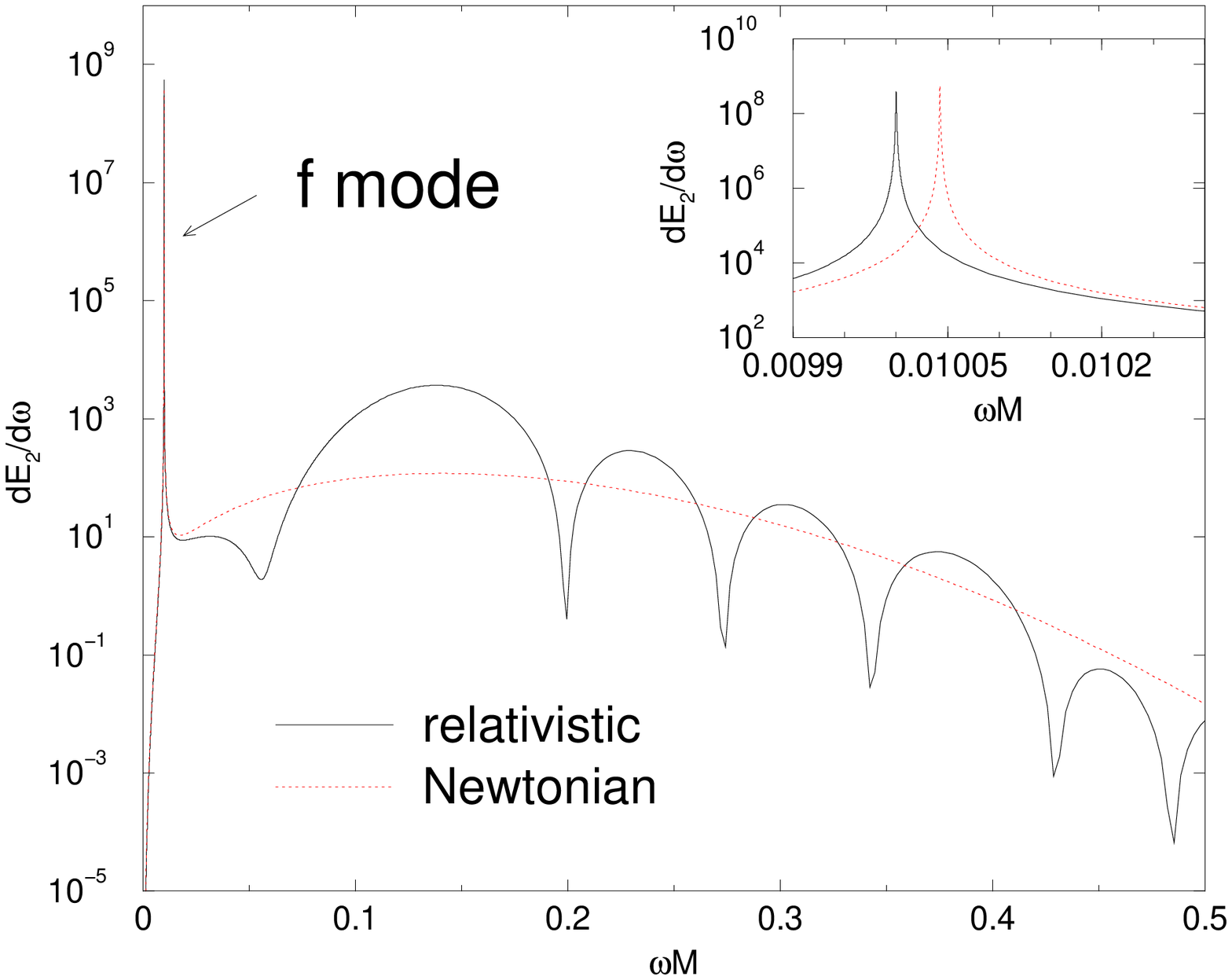}\hspace*{.1\textwidth} \epsfxsize=0.35\textwidth
\epsfbox{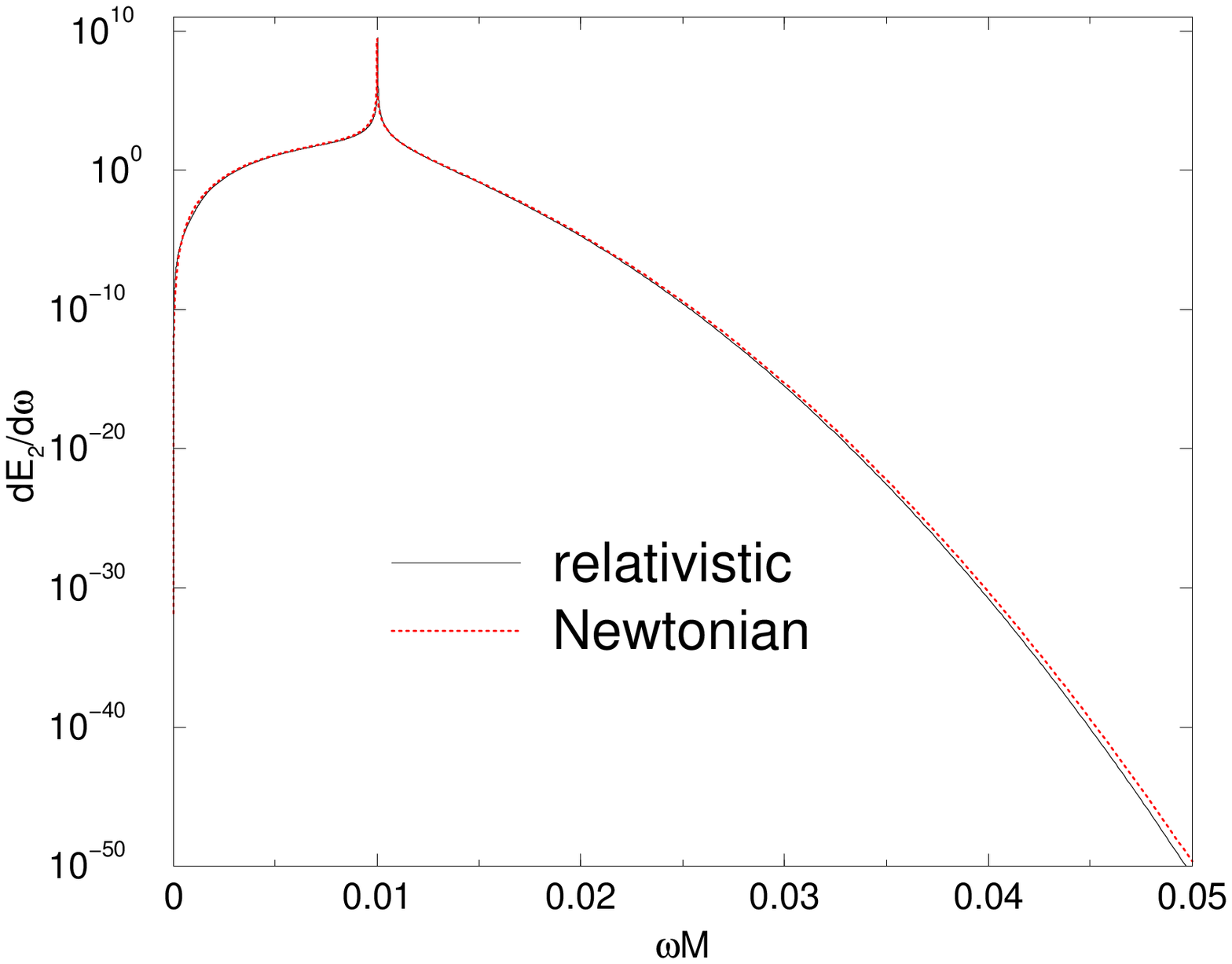} 
\caption{\label{HSR=20M} 
A comparison of Newtonian and relativistic star-only computations for
a weakly relativistic HIF model. In both figures the stellar radius is
$R=20M$, and the shell radius is $R_{\rm shell}=22M$. In (a) the
Gaussian parameter is $a=0.01M^{-2}$, so the shell is not a ``slow
motion'' source. In (b) the Gaussian parameter is
$a=0.00001M^{-2}$, so that the shell is a slow motion source, and the
Newtonian and relativistic results agree in all features.  The $f$
mode frequency for this model is $(0.01+i0.32) 10^{-6}M^{-1}$ in
Newtonian theory and approximately the same in relativistic
computations.  }
\end{figure}

In order to illustrate how the methods of this paper work for a very
weakly relativistic model, Fig.~\ref{HSR=20M} shows two spectra for a
HIF star with $R=20M$. In both, the Newtonian star-only spectrum is
compared with the relativistic star-only (i.e.\,, subtracted)
spectrum, and for both $R_{\rm shell}=22M$.  Figure~\ref{HSR=20M}(a)
shows the case for a Gaussian parameter $a=0.01M^{-2}$, while (b)
shows the case $a=0.00001M^{-2}$. The Newtonian and relativistic
computations of $f$ mode excitation (shown in the inset) agree quite
well in Fig.~\ref{HSR=20M}(a), and there is general agreement in the
overall shape of the spectrum but, as in Fig.~\ref{HSR=5M}(b), the
relativistic case has structure that is missing in the Newtonian
case. This is due to the fact that the shell stress energy source has
a timescale ($\sim a^{-1/2}=10M$) that is less than the light travel
time across the star, and the star is not radiating in phase. By
comparison, in Fig.~\ref{HSR=20M}(b) the shell timescale is $\sim
a^{-1/2}\sim300M$ and the slow approximation is justified. (Note that
the $f$ mode oscillations, at $\omega\sim0.01M^{-1}$ are also slow
compared to the light travel time across the star.)  Because of this,
the Newtonian and relativistic spectra in Fig.~\ref{HSR=20M}(b) are in
excellent overall agreement.



\section{Conclusions}\label{sec:conc} 


We have presented a method of probing the gravitational wave
properties of neutron stars by using time varying stress energy in a
spherical shell. In particular, we have shown that this method can
give more useful answers about the neutron star physics than those
given by studies of the scattering of gravitational waves\cite{AK1} or
by the response of the star to a close particle
orbit\cite{FGB,TSM,ZP}.  The main motivation for considering such a
probe is to compare relativistic and Newtonian computations of
gravitational radiation.  We have shown that the shell probe is well
suited for this purpose, since both the Newtonian and relativistic
computations can be carried out for the model.


Two additional features of the shell probe have been shown to be
important or useful.  One is the possibility of approximately
distinguishing the radiation that can be ascribed to the star from the
radiation due to the shell. Since the radiation from the shell is
stronger than that from the the stellar fluid, this separation is
valuable in bringing out the physical radiation that is of primary
interest.

A fundamentally important feature of the shell probe model, is 
the ability to choose the timescale of the shell
stress energy. This has allowed us to direct attention to the fact that
the Newtonian approximation is not only a weak field approximation,
but a ``slow'' approximation. That is, the quadrupole approximation
used in a quasi-Newtonian gravitational wave calculation supposes that
the light travel time across the source is much smaller that the
period of the waves generated. We have shown that even for a
weakly relativistic stellar model there is not good
agreement in the details of the Newtonian and relativistic
computations if the timescale for the shell stress energy is short.

The results shown in the previous section make this clear.  The
structure and dynamics of the $R=20M$ model of Fig.~\ref{HSR=20M}(a)
is well described by Newtonian physics, but there are large
differences between the Newtonian and relativistic spectra when the
star is excited by a short time scale perturbation.
The difference between the Newtonian and relativistic results is even 
larger for the $R=5M$ model of Fig.~\ref{HSR=5M}(b). 
For smaller Gaussian parameters $a$ (i.e.\,, for driving perturbations
with longer time scales) the ``relativistic-only'' structure seen in 
Figs.~\ref{HSR=5M}(b) and \ref{HSR=20M}(a) decreases.

What becomes clear from these results is that the question of whether
Newtonian physics is adequate for neutron star dynamics is inseparable
from the question of the timescale of the excitation of the neutron
star.  If the timescale is imposed from a distance many times the
neutron stars radius, then the excitation will be slow and our results
(based on a very limited exploration of models) suggest that Newtonian
physics will suffice. On the other hand, rapid processes, due to
impacts, collapse, etc., may be have a timescale only  several times
$GM/c^{3}$, the slow approximation may be violated, and Newtonian
calculations may be significantly in error.

As explained in Sec.\,I, the motivation for the shell probe for
neutron star oscillations is a sequence of two questions.  In
Ref.\,\cite{AK1} Andersson and Kokkotas questioned whether Newtonian
physics was adequate for neutron star physics. The second question is
whether this can be adequately studied with the particle-scattering
model\cite{FGB,TSM,ZP} and its inflexible timescales.  The results
presented here suggest that in cases in which neutron stars are
excited on a very short time scale, those particle-scattering
computations are {\em not} a sufficient basis for the conclusion that
relativistic effects are unimportant in neutron star models of
gravitational radiation sources.

Our main purpose here has been to introduce the shell probe, and the
motivation for it. We have applied it only to a single simple neutron
star model. It is quite possible, of course, that for some equations
of state the importance of relativistic effects might be quite
different. If other equations of state are to be studied towards this
end, we suggest that the method of time varying stress energy in a
shell be considered as a good way of getting the clearest comparison
of Newtonian and relativistc predictions.

\section{Acknowledgments}
We thank Nils Andersson for supplying the frequencies of the even
parity modes of homogeneous stars. This work was partially supported 
by NSF grant PHY9734871. Z. A. was supported by PRAXIS
XXI/BD/3305/94 grant from FCT (Portugal).


\appendix
\section{The newtonian limit of a HIF}

\subsection{The newtonian perturbation equations and their solution}
We consider a HIF, excited by a spherical shell whose mass is much
smaller than the star's mass. Assuming that the motion of the fluid
consists of small perturbations around a spherically symmetric and
static fluid ball (the equilibrium star), we can decompose all
perturbative scalar quantities, namely the gravitational potential,
the pressure, and the density perturbations, in scalar spherical
harmonics, and can decompose the fluid's velocity in even parity
vector spherical harmonics. We write these decompositions as
\begin{eqnarray}
& &\label{vv1}
U=U_{eq}(r)+\sum_{l}\delta U_{l}(r,t)Y_{l0}(\theta)\\
& &\label{vv2}
p=p_{eq}(r)+\rho_{eq}\sum_{l}\delta h_{l}(r,t)Y_{l0}(\theta)\\
& &\label{density}
\rho=\rho_{eq}+\sum_{l}\left\{\frac{e^{-at^{2}}}{M}\delta[r-R_{\rm shell}]-
\frac{W_{l}(r,t)}{r^{2}}\rho_{eq}\delta[r-R]\right\}Y_{l0}(\theta)\\
& &\label{vv3}
v^{r}=-\frac{1}{r^{2}}\sum_{l}
\frac{\partial}{\partial t}W_{l0}(r,t)Y_{l0}(\theta)\\
& &\label{vv4}
v^{\theta}=\frac{1}{r^{2}}\sum_{l}\frac{\partial}{\partial t}V_{l0}(r,t)
\frac{\partial}{\partial\theta}Y_{l0}(\theta)\ ,
\end{eqnarray}
where $U_{eq}, p_{eq}, \rho_{eq}=\rm const$ are the gravitational
potential, pressure and constant density of the spherical
(equilibrium) star. The perturbation of the density has two
contributions: one from the matter in the shell and the other from the
star itself. In writing the first one, we supposed that $\delta
T_{00,\rm shell}\approx \delta\rho_{\rm shell}$ is dominant over all
the other components of the stress energy tensor of the shell (weak
field, slow motion approximations) and that $\delta T_{00,\rm shell}$
is given by Eqs.\,(\ref{shellstress}) and (\ref{gaussian}). The
density due to the fluid perturbation is zero everywhere inside the
star (since the star is incompressible), but not at the unperturbed
surface of the star\cite{WILL}. In writing the perturbation of the
pressure, we used the definition Eq.\,(\ref{enthalpy}) and the fact
that $p_{eq}\ll \rho_{eq}$ in the weak field limit.

We then substitute these expressions in the fluid equations for a
perfect barotropic fluid, which are Poisson's equation for the
gravitational potential,
\begin{equation}\label{pt1}
\nabla^{2}U=-4\pi \rho\ ,
\end{equation}
the continuity equation,
\begin{equation}\label{pt2}
\rho_{,t}=-\vec{\nabla}.(\rho\vec{v})\ ,
\end{equation}
and Euler's equation,
\begin{equation}\label{pt3}
\vec{v}_{,t}+(\vec{v}.\vec{\nabla})\vec{v}=\vec{\nabla}U-\frac{1}{\rho}
\vec{\nabla}p\ ,
\end{equation}
and we keep terms only to first order in the perturbations. The
resulting linearized fluid equations can then be reduced to two
equations: one for $\delta U_{l}$ and another for $\delta
h_{l}$. These equations are the Newtonian limit of the relativistic
even parity perturbation equations derived by Lindblom {\it et
al}.\cite{LMI}.  In the special case of a HIF, they reduce to two
decoupled second order equations,
\begin{equation}\label{potenteq}
\delta U_{l,rr}+\frac{2}{r}\delta U_{l,r}-\frac{l(l+1)}{r^{2}}\delta U_{l}=
-4\pi\left\{\frac{e^{-at^{2}}}{M}\delta[r-R_{\rm shell}]-
\frac{W_{l0}(r,t)}{r^{2}}\rho_{eq}\delta[r-R]\right\}
\end{equation}
and 
\begin{equation}\label{enthalpeq}
\delta h_{l,rr}+\frac{2}{r}\delta h_{l,r}
-\frac{l(l+1)}{r^{2}}\delta h_{l}=0\ .
\end{equation}
Since the left hand sides of Eqs.\,(\ref{potenteq}) and
(\ref{enthalpeq}) are equivalent to the multipole decomposition of the
Laplacian, it is straightforward to write down the solutions that are
well behaved at the center of the star and that vanish at infinity:
\begin{equation}
\delta U_{l}(r,t)=\left\{\begin{array}{ll}
\alpha_{l}(t)r^{l}+\frac{4\pi}{2l+1}
\frac{r^{l}}{MR^{l-1}_{\rm shell}}e^{-at^{2}},
& \quad r\leq R\\
\: & \:\\
\frac{\beta_{l}}{r^{l+1}}+\frac{4\pi}{2l+1}
\frac{r^{l}}{MR^{l-1}_{\rm shell}}
e^{-at^{2}}, &\quad R\leq r\leq R_{\rm shell}\\
\: & \: \\
\frac{\beta_{l}}{r^{l+1}}+\frac{4\pi}{2l+1}
\frac{R^{l+2}_{\rm shell}}{Mr^{l+1}}
e^{-at^{2}}, & \quad r\geq R_{\rm shell}
\end{array}\right.
\end{equation}
and 
\begin{equation}\begin{array}{ll}
\delta h_{l}(r,t)=\mu_{l}(t)r^{l}, & \quad r<R\ .
\end{array}
\end{equation}
The three constants can be easily determined, by requiring the
vanishing of the Lagrangian pressure at $r=R$, as in Eq.\,(\ref{pressure1}), 
by requiring 
the
continuity of $\delta U_{l}$ at the surface of the star, and by integrating 
Eq.\,(\ref{potenteq}) about $r=R$ and by using the linearized Euler equation
for the radial component of the fluid's velocity.
The result is
\begin{mathletters}
\begin{eqnarray}\label{thewhole}
& &
\beta_{l}=R^{2l+1}\alpha_{l}\\
& &
\mu_{l}=(2l+1)\frac{\alpha_{l}}{3}\\
& &\label{harmonic}
\frac{d^{2}\alpha_{l}}{dt^{2}}+\omega^{2}_{l}\alpha_{l}=
\frac{16\pi^{2}}{(2l+1)^{2}MR^{l-1}_{\rm shell}}l\rho_{eq}e^{-at^{2}}\\
& &
\omega^{2}_{l}=\frac{8\pi \rho_{eq}}{3}\frac{l(l-1)}{2l+1}\ .
\end{eqnarray}
\end{mathletters}
The frequency $\omega_{l}$ is the $f$ mode of vibration of the star.

\subsection{Damping of the $f$ mode oscillation}
Once set into vibration at its $f$ mode frequency by the shell
perturbation, the star would oscillate forever, since there is no
damping mechanism for the perfect fluid in Newtonian theory. This is
clear from Eq.\,(\ref{harmonic}), which is the equation of an undamped
harmonic oscillator. But in practice the star will radiate away its
vibrational energy through gravitational waves, over a long period of
time. The time, $\tau_{l}$, for gravitational wave damping of the $f$
mode oscillation can be computed, in the weak field, slow motion,
approximation using energy conservation\cite{KIP2}\cite{DETWEILER} to
be,
\begin{equation}
\tau_{l}=\frac{4l(l-1)^{2}(2l+1)[(2l-1)!!]^{2}}{3(l+1)(l+2)\omega^{2l+2}_{l}
R^{2l+1}}\ .
\end{equation}
To introduce this damping we replace Eq.\,(\ref{harmonic}) by
\begin{equation}
\frac{d^{2}\alpha_{l}}{dt^{2}}+\omega^{2}_{l}\alpha_{l}+
\frac{2}{\tau_{l}}\frac{d\alpha_{l}}{dt}=
\frac{16\pi^{2}}{(2l+1)^{2}MR^{l-1}_{\rm shell}}l\rho_{eq}e^{-at^{2}}\ .
\end{equation}
(For a similar procedure see \cite{TURNER}
and \cite{WILL}).
The retarded solution of this equation is
\begin{equation}\label{alphaeq}
\alpha_{l}(t)=\frac{16\pi^{2}l\rho_{eq}}{(2l+1)^{2}MR^{l-1}_{\rm shell}
\omega_{l}}\int_{0}^{\infty}dv e^{-v/\tau_{l}}e^{-a[t-v]^{2}}
\sin[\omega_{l}v]\ .
\end{equation}
From Eq.\,(\ref{thewhole}) and the  linearized Euler equation
we obtain
\begin{equation}
\frac{W_{l}(r,t)}{R^{2}}\rho_{eq}=\frac{2l+1}{4\pi}R^{l-1}\alpha_{l}\ .
\end{equation}
Combined with Eq.\,(\ref{alphaeq}) and Eq.\,(\ref{density}) for $l=2$,
this leads to the following expression for the quadrupole perturbation
of the density
\begin{equation}\label{quaddens}
\delta\rho_{2}(r,t)=\frac{e^{-at^{2}}}{M}\delta[r-R_{\rm shell}]+
\frac{8\pi R\rho_{eq}}{5\omega_{2}MR_{\rm shell}}\int_{0}^{\infty}
dv e^{-v/\tau_{2}}e^{-a[t-v]^{2}}\sin[\omega_{2}v]\delta[r-R]\ .
\end{equation}

\subsection{The energy radiated by a vibrating, semi-Newtonian HIF}
The energy radiated in gravitational waves by an oscillating Newtonian
star can be computed by regarding the star's gravitational field as a
small perturbation of Minkowski spacetime. For details see
Ref.\,\cite{KIP}.  The energy radiated in the quadrupole is
\begin{equation}\label{newtspec}
\frac{dE_{2}}{d\omega}=\frac{1}{32\pi^{2}}\omega^{6}|A_{20}(\omega)|^{2}
\end{equation}
where $A_{20}$ is the Fourier amplitude of the mass  quadrupole,
\begin{equation}\label{quadamp}
A_{20}(\omega)=\int_{-\infty}^{\infty}du e^{i\omega u}I_{20}(u)\ .
\end{equation}
In the slow motion approximation, $I_{20}(u)$ is simply
\begin{equation}\label{quadmom} 
I_{20}(u)=\frac{16\pi}{5\sqrt{3}}\int_{0}^{\infty}\delta\rho_{2}(u,r)
r^{4}dr \ .
\end{equation}
In this approximation, Eqs.\,(\ref{quaddens}) and (\ref{quadamp}) can
be combined to give the Fourier amplitude
\begin{equation}\label{finalexp}
A_{20}(\omega)=\frac{16\pi}{5\sqrt{3}M}\sqrt{\frac{\pi}{a}}
e^{-\omega^{2}/(4a)}\left\{R^{4}_{\rm shell}+
\frac{8\pi\rho_{eq}}{5}\frac{R^{5}}{R_{\rm shell}}
\frac{[\omega^{2}_{2}-\omega^{2}+2i\omega/\tau_{2}]}{(\omega^{2}_{2}-
\omega^{2})^{2}+4\omega^{2}/\tau^{2}_{2}}\right\}\ ,
\end{equation}
which can then be used in Eq.\,(\ref{newtspec}) to compute the
Newtonian quadrupole energy spectrum.  The first term in curly
brackets in the Fourier amplitude Eq.\,(\ref{finalexp}) is the
contribution only of the shell to the total energy radiated. The
second term is the contribution only of the star's oscillations to the
energy. This should be contrasted with the relativistic procedure in
which no such exact identification of the shell and star contributions
can be made.

\appendix



\end{document}